\documentclass[twocolumn,showpacs,preprintnumbers]{revtex4}
\usepackage{graphicx}
\usepackage{amssymb}

\begin{document}
\title{The light cone at infinity}
\author{Sean A. Hayward}
\affiliation{Department of Science Education, Ewha Womans University,
Seodaemun-gu, Seoul 120-750, Korea\\ {\tt hayward@mm.ewha.ac.kr}}
\date{29th December 2003}

\begin{abstract}
A recent refinement of Penrose's conformal framework for asymptotically flat
space-times is summarized. The key idea concerns advanced and retarded
conformal factors, which allow a rigid description of infinity as a locally
metric light cone. In the new framework, the Bondi-Sachs energy-flux integrals
of ingoing and outgoing gravitational radiation decay at spatial infinity such
that the total radiated energy is finite, and the Bondi-Sachs energy-momentum
has a unique limit at spatial infinity, coinciding with the uniquely rendered
ADM energy-momentum.
\end{abstract}
 \maketitle

\section{Introduction}
Space-time asymptotics, the study of isolated gravitational systems at large
distances, is important in General Relativity because it allows exact
definitions of physical quantities which are otherwise not known, such as the
mass-energy of a system and the energy flux of gravitational radiation, related
by the Bondi-Sachs energy-loss equation \cite{B,BBM,S}. Penrose \cite{P,PR}
formulated this using conformal transformations, so that infinite distances and
times are rendered finite: infinity becomes a mathematically finite boundary
where exact formulas can be derived. This is now standard textbook theory
\cite{PR,W}, yet fundamental issues remain. Penrose's theory describes future
null infinity $\Im^+$, whereas spatial infinity $i^0$ has a different theory
\cite{ADM,G}. In particular, the Bondi-Sachs energy (at $\Im^+$) does not
necessarily reduce appropriately to the ADM energy (at $i^0$) \cite{AH,AM,A}. A
surprisingly unresolved question is whether initial data on an asymptotically
flat spatial hypersurface (or past null infinity $\Im^-$) determines final data
at $\Im^+$.

This presentation summarizes a new approach, described in detail recently
\cite{inf}. The key idea concerns advanced and retarded conformal factors,
which do the work of the Penrose conformal factor at $\Im^+$ and $\Im^-$
respectively, while enforcing appropriate structure at $i^0$. The factors
differentially relate physical and conformal null coordinates at infinity.

\section{Example}
\begin{figure}
\includegraphics[width=5cm,height=7cm,angle=0]{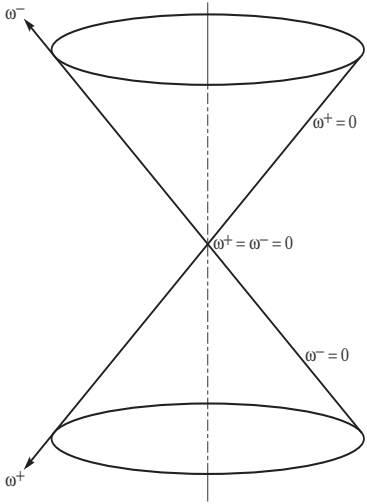}
 \caption{The light cone at infinity, $\Im$, depicting the advanced and retarded
 conformal factors $\omega^\pm$. The physical space-time ($\omega^+>0$,
 $\omega^->0$) lies outside this asymptotic light cone. Radiation propagates in
 from past null infinity $\Im^-$ ($\omega^-=0$) and out to future null infinity
 $\Im^+$ ($\omega^+=0$). Spatial infinity $i^0$ ($\omega^+=\omega^-=0$) is the
 vertex.}
 \label{fig}
\end{figure}

As a simple example, the Schwarzschild space-time with mass $M$ in standard
coordinates is given by
\begin{equation}
ds^2=r^2dS^2+(1-2M/r)^{-1}dr^2-(1-2M/r)dt^2
\end{equation}
where $dS^2$ refers to the unit sphere. This can be written in dual-null
coordinates $2\xi^\pm=t\pm r_*$, where $r_*=r+2M\ln(r-2M)$, as
\begin{equation}
ds^2=r^2dS^2-(1-2M/r)4d\xi^+d\xi^-.
\end{equation}
The physical null coordinates $\xi^\pm$ are now transformed to conformal null
coordinates $\psi^\pm$ by, for instance, inversion $\xi^\pm=-1/\psi^\pm$, so
that $\xi^\pm\to\pm\infty$ are rendered finite, $\psi^\pm=0$. The null
coordinates are differentially related by
\begin{equation}
\omega^\pm=\left(\frac{d\psi^\pm}{d\xi^\pm}\right)^{1/2}
\end{equation}
which are $\mp\psi^\pm$ in this case. Since the physical metric $g$ has an
angular part $r^2dS^2$ which becomes infinite as $r\to\infty$, Penrose's idea
was to regularize it by transforming to a conformal metric $\Omega^2g$, where
$\Omega\sim1/r$ is the conformal factor. Simultaneously, the normal part
$-(1-2M/r)4d\xi^+d\xi^-$, which was already regular, is being multiplied by
$\Omega^2$ and being transformed by
$d\xi^+d\xi^-=d\psi^+d\psi^-/(\omega^+\omega^-)^2$. The obvious way to keep it
regular is to choose
\begin{equation}\label{Omega}
\Omega=\omega^+\omega^-.
\end{equation}
Then
\begin{eqnarray}
\Omega^2ds^2&=&(\psi^+\psi^-r)^2dS^2-(1-2M/r)4d\psi^+d\psi^-\\ &=&(\rho
r/r_*)^2dS^2+(1-2M/r)(d\rho^2-d\tau^2)\\
&\to&\rho^2dS^2+d\rho^2-d\tau^2\qquad\hbox{as $r\to\infty$}
\end{eqnarray}
where
\begin{equation}\label{rhotau}
2\psi^\pm=\tau\mp\rho.
\end{equation}
The conformal metric is regular, actually becoming flat, as $r\to\infty$
($r_*/r\to1$). Physical infinity has become a light cone $\Omega=0$, the light
cone at infinity \cite{AH}, as depicted in Fig.~\ref{fig}. Spatial infinity
$i^0$ becomes a point $\omega^+=\omega^-=0$ and the physical space-time
$\omega^\pm>0$ lies outside its light cone. Null infinity $\Im^\pm$ is given by
$\omega^\pm=0$, $\omega^\mp\not=0$.

Clearly there is more information in the sub-factors $\omega^\pm$ than in the
Penrose factor $\Omega$ alone. This structure can be used to refine Penrose's
definition of asymptotic flatness at $\Im^\pm$ to include $i^0$. For obvious
reasons, $\omega^+$ and $\omega^-$ are called the advanced and retarded
conformal factors respectively \cite{inf}. While Penrose emphasized the metric
transformation, now the dual-null coordinate transformation is to be considered
in conjunction.

\section{Definition}
The above ideas motivated the following definition \cite{inf}. A space-time
$(M,g)$ is asymptotically flat if: (i) there exists a space-time $(\hat M,\hat
g)$ with boundary $\Im=\partial\hat M$ and functions $\omega^\pm>0$ on $M$ such
that $\hat M=M\cup\Im$ and $\hat g=(\omega^+\omega^-)^2g$ in $M$; (ii) $\Im$ is
locally a light cone with respect to $\hat g$, with vertex $i^0$ and future and
past open cones $\Im^+$ and $\Im^-$ respectively; (iii) $\omega^\pm=0$,
$\omega^\mp\not=0$, $\hat\nabla\omega^\pm\not=0$ on $\Im^\pm$, and $\hat
g^{-1}(\hat\nabla\omega^+,\hat\nabla\omega^-)\not=0$ on $\Im$; (iv) ($\hat
g,\omega^\pm$) are $C^k$ on $\hat M-i^0$, e.g.\ $k=3$. Differentiability at
$i^0$ is more complex, as described subsequently.

The definition recovers the basic conditions of Penrose's definition of
asymptotic simplicity \cite{P,PR}, namely $\Omega=0$ and
$\hat\nabla\Omega\not=0$ on $\Im^\pm$, with $\Omega$ as before (\ref{Omega}).
Also $\omega^+=\omega^-=0$ at $i^0$ as desired. This leaves considerable gauge
freedom in the conformal factors $\omega^\pm$, namely
$\omega^\pm\mapsto\alpha_\pm\omega^\pm$ for any $C^k$ functions $\alpha_\pm>0$
on $\hat M-i^0$. This allows $\hat g$ to be fixed so that $\Im$ is locally a
metric light cone \cite{inf}:
\begin{eqnarray}
d\hat s^2&\approx&\rho^2dS^2+d\rho^2-d\tau^2\\
&=&(\psi^--\psi^+)^2dS^2-4d\psi^+d\psi^-
\end{eqnarray}
where $\approx$ denotes equality on $\Im$ in a neighbourhood of $i^0$. The
gauge conditions are
\begin{eqnarray}\label{gauge}
\hat g^{-1}(\hat\nabla\omega^\pm,\hat\nabla\omega^\pm)/\omega^\pm&\approx&0\\
\hat g^{-1}(\hat\nabla\omega^+,\hat\nabla\omega^-)&\approx&1/2
\end{eqnarray}
and they imply
\begin{eqnarray}
\hat g^{-1}(\hat\nabla\Omega,\hat\nabla\Omega)/\Omega&\approx&1\\
2\hat\nabla\otimes\hat\nabla\Omega&\approx&\hat g.
\end{eqnarray}
The last two expressions vanish in the usual gauge chosen for $\Im^+$
\cite{G,W}, in which it is a metric cylinder. In the new gauge,
\begin{equation}
\psi^\pm\approx\mp\omega^\pm
\end{equation}
as in the Schwarzschild case.

\section{Implementation}
The new framework has been implemented using the spin-coefficient or
null-tetrad formalism \cite{PR}. The basic objects are a spin-metric
$\varepsilon$ and a spin-basis $(o,\iota)$, or a metric
$g=\varepsilon\circ\bar\varepsilon$ and a null tetrad $(l,m,\bar m,l')=(o\bar
o,o\bar\iota,\iota\bar o,\iota\bar\iota)$. The key result \cite{inf} is that
the desired conformal transformations are given by
\begin{eqnarray}\label{spin}
\varepsilon&=&\hat\varepsilon/\omega\omega'\\ (o,\iota)&=&(\omega\hat
o,\omega'\hat\iota)
\end{eqnarray}
or
\begin{eqnarray}\label{metric}
g&=&\hat g/(\omega\omega')^2\\ (l,m,\bar m,l')&=&(\omega^2\hat
l,\omega\omega'\hat m,\omega\omega'\hat{\bar m},\omega'^2\hat l')
\end{eqnarray}
where $\omega$ and $\omega'$ now denote respectively the advanced and retarded
conformal factors.

The physical and conformal tetrad derivative operators are related by
\begin{equation}\label{deriv}
(D,\delta,\delta',D')=(\omega^2\hat
D,\omega\omega'\hat\delta,\omega\omega'\hat\delta',\omega'^2\hat D')
\end{equation}
and the weighted spin-coefficients are related by
\begin{eqnarray}\label{sc0}
\kappa&=&\hat\kappa\omega^3/\omega'\\ \sigma&=&\omega^2\hat\sigma\\
\rho&=&\omega^2(\hat\rho+\hat D\log\omega\omega')\\
\tau&=&\omega\omega'(\hat\tau+\hat\delta\log\omega\omega')\\
\tau'&=&\omega\omega'(\hat\tau'+\hat\delta'\log\omega\omega')\\
\rho'&=&\omega'^2(\hat\rho'+\hat D'\log\omega\omega')\\
\sigma'&=&\omega'^2\hat\sigma'\\\label{sc1}
\kappa'&=&\hat\kappa'\omega'^3/\omega.
\end{eqnarray}
These formulae are all inverses of those in \cite{inf}.

\section{Regularity}
Asymptotic expansions valid near the whole of $\Im$, including $i^0$, can now
be developed. There is a canonical coordinate system near $\Im$, obtained by
propagating the metric spheres at $\Im$ into the physical space-time along null
hypersurfaces, generating a dual-null foliation of transverse spatial surfaces.
The null coordinates $(\psi,\psi')=\sqrt2(\psi^+,\psi^-)$ are related by
$g(\hat l')=d\psi$, $g(\hat l)=d\psi'$ and the dual-null gauge entails certain
relations between the spin-coefficients \cite{bhs}, including
$\kappa=\kappa'=0$. The conformal freedom can be fixed as before by
\begin{equation}
(\omega,\omega')=(-\psi,\psi')/\sqrt2.
\end{equation}
It is also convenient to introduce a hyperboloidal spatial function
\begin{equation}
u=\frac{\omega\omega'}{\omega+\omega'}
\end{equation}
which behaves as $u\sim1/r$ at $\Im$, where $a\sim b$ means $a/b\approx1$ and
the radial function $r$ relates the area form ${*}1=\tilde{*}r^2$ of the
transverse surfaces to the area form $\tilde{*}1$ of a unit sphere. With the
asymptotic gauge choice
\begin{equation}
\chi\approx1
\end{equation}
the conformal expansions are determined for the metric light cone as
\begin{eqnarray}\label{rho0}
\hat\rho&\sim&1/\sqrt2(\omega+\omega')\\
\label{rho1} \hat\rho'&\sim&-1/\sqrt2(\omega+\omega').
\end{eqnarray}
Using the spin-coefficient transformations (\ref{sc0})--(\ref{sc1}), the
physical expansions are found to be
\begin{eqnarray}
\rho&\sim&-u/\sqrt2\\ \rho'&\sim&u/\sqrt2.
\end{eqnarray}
Now it is demanded that $\Im$ be a smoothly embedded metric light cone. This
leads to the following asymptotic behaviour for the remaining conformal
weighted spin coefficients:
\begin{eqnarray}
\hat\sigma&=&O(u/\omega)\\ \hat\sigma'&=&O(u/\omega')\\
\hat\tau&=&O(\omega\omega')\\ \hat\tau'&=&O(\omega\omega')
\end{eqnarray}
where $a=O(b)$ here means that $a/b$ is regular at $\Im$, where a function
$f(\omega,\omega',\theta,\phi)$ is said to be regular at $\Im$ if the following
limits exist:
\begin{eqnarray}
f_+&=&\lim_{\omega\to0}f\quad\hbox{at $\Im^+$}\\
f_-&=&\lim_{\omega'\to0}f\quad\hbox{at $\Im^-$}\\ f_0&=&\lim_{\omega\to0}f_-
=\lim_{\omega'\to0}f_+\quad\hbox{at $i^0$.}
\end{eqnarray}
Note that this allows angular dependence $f_0(\theta,\phi)$ at $i^0$, but not
boost dependence. Thus for some purposes it may be useful to expand $i^0$ from
a point to a sphere. In terms of the physical spin-coefficients, this yields
\begin{eqnarray}\label{shear0}
\sigma&=&O(u\omega)\\\label{shear1} \sigma'&=&O(u\omega')\\\label{twist0}
\tau&=&O((\omega\omega')^2)\\\label{twist1} \tau'&=&O((\omega\omega')^2).
\end{eqnarray}
These are geometrically motivated conditions, yet they seem to describe a class
of space-times with desired physical properties, as follows.

\section{Energy}
The Bondi-Sachs energy \cite{B,BBM,S} at $\Im^+$ can be expressed as
\cite{P,PR,NT,mon}
\begin{equation}
E_{BS}=\lim_{\omega\to0}\frac{A^{1/2}}{(4\pi)^{3/2}}
\oint{*}\frac{\sigma\sigma'-\Psi_2}{\chi\bar\chi}
\end{equation}
where $A=\oint{*}1$ is the area of a transverse surface and integrals $\oint$
refer to transverse surfaces. Similarly, the ADM energy \cite{ADM} at $i^0$ can
be expressed as \cite{G,AH,AM,A,qle}
\begin{equation}
E_{ADM}=-\lim_{\omega\to0}\lim_{\omega'\to0}\frac{A^{1/2}}{(4\pi)^{3/2}}
\oint{*}\frac{\Re\Psi_2}{\chi\bar\chi}
\end{equation}
where, in the original treatment, it was unclear whether the limit depended on
the boost direction $\omega/\omega'$, i.e.\ on the choice of spatial
hypersurface. These two expressions are not obviously consistent, and indeed,
$E_{ADM}$ is not the limit of $E_{BS}$ at $i^0$ without some extra assumption
\cite{AM}. These questions can be addressed using the Hawking quasi-local
mass-energy \cite{H}, which can be written as \cite{mon}
\begin{equation}
E=\frac{A^{1/2}}{(4\pi)^{3/2}}\oint{*}\frac{K+\rho\rho'}{\chi\bar\chi}
=\frac{A^{1/2}}{(4\pi)^{3/2}} \oint{*}\frac{\sigma\sigma'-\Psi_2}{\chi\bar\chi}
\end{equation}
where the second expression follows in vacuo from the definition of complex
curvature $K$. In order for the total energy to be finite, one needs to further
expand
\begin{eqnarray}\label{exp0}
\chi-1&\sim&\chi_1u\\ -\sqrt2\rho/u-1&\sim&\rho_1u\\\label{exp1}
\sqrt2\rho'/u-1&\sim&\rho'_1u
\end{eqnarray}
where $(\chi_1,\rho_1,\rho'_1)$ are regular at $\Im$. Then
\begin{equation}
E_{BS}=\lim_{u\to0}E=\frac1{8\pi}\oint{\tilde*}(\chi_1+\bar\chi_1-\rho_1-\rho'_1).
\end{equation}
In words, $E_{BS}$ exists at $\Im$, is the limit of the Hawking energy, and has
a unique limit at $i^0$ in this framework. In order to compare with the ADM
energy, one can write the leading-order terms in the shears
(\ref{shear0}--\ref{shear1}) as
\begin{eqnarray}
\sigma_1&=&\sigma/u\omega\\ \sigma'_1&=&\sigma'/u\omega'
\end{eqnarray}
which are regular at $\Im$. Then the discrepancy ``$E_{BS}-E_{ADM}$'' (if
$E_{ADM}$ were extended to from $i^0$ to $\Im^\pm$ by the same formula) is
found to be $\oint\tilde{*}\sigma_1\sigma_1'(\omega+\omega')/4\pi$, which is
generally non-zero at $\Im^\pm$ ($\omega=0$ or $\omega'=0$), but vanishes at
$i^0$ ($\omega=\omega'=0$) from any direction. Thus the ADM energy is the limit
of the Bondi-Sachs energy at spatial infinity in this framework, and also the
limit of the Hawking energy from any spatial or null direction:
\begin{equation}
E_{ADM}=\lim_{\omega+\omega'\to0}E_{BS}=\lim_{\omega\to0}\lim_{\omega'\to0}E.
\end{equation}
This resolution explicitly rests on the additional structure at spatial
infinity provided by the advanced and retarded conformal factors
$(\omega,\omega')$.

The asymptotic regularity conditions and expansions
(\ref{shear0}--\ref{twist1},\ref{exp0}--\ref{exp1}) can be used to show,
assuming either vacuum or suitable fall-off of the matter fields, that $(\hat
DE,\hat D'E)$ are $O(1)$, as expected on physical grounds. In particular, in
the vacuum case they imply
\begin{eqnarray}\label{massgain}
\sqrt2DE_{BS}&=&\frac1{4\pi}\oint{\tilde*}|N|^2 \quad\hbox{at $\Im^-$
($\omega'=0$)}\\\label{massloss}
\sqrt2D'E_{BS}&=&-\frac1{4\pi}\oint{\tilde*}|N'|^2 \quad\hbox{at $\Im^+$
($\omega=0$)}
\end{eqnarray}
where
\begin{eqnarray}
N&=&\sigma/u\\ N'&=&\sigma'/u
\end{eqnarray}
are the retarded and advanced Bondi-Sachs news functions. Here (\ref{massloss})
is the usual Bondi-Sachs energy-loss equation, showing that the outgoing
gravitational radiation carries energy away from the system, with energy
density $|N'|^2/r^2$. Similarly (\ref{massgain}) shows that ingoing
gravitational radiation supplies energy to the system. In conformal null
coordinates,
\begin{eqnarray}
\sqrt2\hat DE&=&\frac1{4\pi}\oint{\tilde*}|\sigma_1|^2 \quad\hbox{at $\Im^-$}\\
\sqrt2\hat D'E&=&-\frac1{4\pi}\oint{\tilde*}|\sigma'_1|^2 \quad\hbox{at
$\Im^+$}
\end{eqnarray}
which are regular in the limit at $i^0$. Thus the change in energy from $i^0$
to a section of $\Im^\pm$ is finite. Physically this means that the ingoing and
outgoing gravitational radiation decays near spatial infinity such that its
total energy is finite. It is straightforward to generalize from energy to
energy-momentum \cite{inf}.

\section{Summary and issues}

\noindent$\bullet$ Penrose's conformal framework has been refined using
advanced and retarded conformal factors.

\noindent$\bullet$ A new definition of asymptotic flatness at both spatial and
null infinity has been given.

\noindent$\bullet$ The light cone at infinity can be locally fixed as a metric
light cone.

\noindent$\bullet$ Asymptotic regularity conditions follow from smooth
embedding of the light cone.

\noindent$\bullet$ The ADM energy-momentum (at $i^0$) is rendered unique.

\noindent$\bullet$ The Bondi-Sachs energy-momentum (at $\Im^\pm$) is extended
to $i^0$, uniquely and consistently with ADM.

\noindent$\bullet$ The energy flux of gravitational radiation (or news) decays
at $i^0$ such that its total energy is finite.

\noindent$\bullet$ A practical implementation using the spin-coefficient or
null-tetrad formalism has been given.

\noindent$\bullet$ The rigid universal structure allows simple behaviour of
physical fields at $\Im$, e.g.\ $(\chi_1,\rho_1,\rho'_1,\sigma_1,\sigma'_1)$
characterize energy-momentum.

\noindent$\bullet$ Higher order asymptotic expansions?

\noindent$\bullet$ Angular momentum, multipole moments?

\noindent$\bullet$ Asymptotic symmetry group? BMS group (at $\Im^\pm$)
simplified by vertex conditions?

\noindent$\bullet$ Tensorial formulation? e.g.\ gravitational radiation in
transverse traceless form.

\noindent$\bullet$ Conformal 3+1 form in $(\tau,\rho,\theta,\phi)$ coordinates?
Regular field equations at $\Im$?

\medskip\noindent Supported by research grant ``Black holes and gravitational
waves'' of Ewha Womans University.

\end{document}